# Picosecond ultrasonics with a free-running dual-comb laser

JUSTINAS PUPEIKIS,[1,*] BENJAMIN WILLENBERG,[1] FRANCOIS BRUNO,[2] MIKE HETTICH,[3] ALEXANDER NUSSBAUM-LAPPING,[1] MATTHIAS GOLLING,[1] CAROLIN P. BAUER,[1] SANDRO L. CAMENZIND,[1] ABDELMJID BENAYAD,[4] PATRICE CAMY,[4] BERTRAND AUDOIN,[2] CHRISTOPHER R. PHILLIPS,[1] AND URSULA KELLER[1]

[1]*Department of Physics, Institute for Quantum Electronics, ETH Zurich, Switzerland*
[2]*Univ. Bordeaux, CNRS, I2M, UMR 5295, F-33400 Talence, France*
[3]*Research Center for Non-Destructive Testing GmbH, 4040 Linz, Austria*
[4]*Centre de Recherche sur Les Ions, Les Matériaux et La Photonique (CIMAP), UMR 6252 CEA-CNRS-ENSICAEN, Université de Caen Normandie, 6 Boulevard Du Maréchal Juin, 14050, Caen Cedex 4, France*
*\*pupeikis@phys.ethz.ch*

**Abstract:** We present a free-running 80-MHz dual-comb polarization-multiplexed solid-state laser which delivers 1.8 W of average power with 110-fs pulse duration per comb. With a high-sensitivity pump-probe setup, we apply this free-running dual-comb laser to picosecond ultrasonic measurements. The ultrasonic signatures in a semiconductor multi-quantum-well structure originating from the quantum wells and superlattice regions are revealed and discussed. We further demonstrate ultrasonic measurements on a thin-film metalized sample and compare these measurements to ones obtained with a pair of locked femtosecond lasers. Our data show that a free-running dual-comb laser is well-suited for picosecond ultrasonic measurements and thus it offers a significant reduction in complexity and cost for this widely adopted non-destructive testing technique.

## 1. Introduction

Picosecond ultrasonics is a technique, developed in the mid-1980s, in which ultrashort laser pulses are used to generate and detect acoustic waves with very short wavelength, typically in a nanometer range [1]. Established applications of the method include the determination of elastic parameters [2–9], acoustic damping properties [10–15], structural properties [16–19], interface adhesion and coupling [20–23] and imaging of embedded layers [24–29]. Ultrafast acoustic dynamics were studied in metals [6,21,30–35], semiconductors [10,36–39], dielectric materials [11,14,16,40] and polymers [41–44]. Recent developments have also shown the feasibility to apply the technique to newly emerging advanced materials [23,45,46]. A wide variety of systems can be investigated ranging from thin-films [3,47,48] to multilayer structures [36,37,46,49–52] and micro nanostructures [32,53–60]. The first developed application area of picosecond ultrasonics was in the microelectronics industry, where accurate characterization of material thickness and bonding layers at a nanometer scale are required [61]. More recently, the capability of the technique to measure mechanical cell properties in a contactless manner was demonstrated [62]. Advances in the measurement methods have allowed picosecond ultrasonics to be applied to the label-free imaging of cells by using the mechanical properties as the contrast modality [63–65]. This imaging capability offers sub-optical in-depth resolution and an in-plane resolution limited by optical diffraction. Here listed applications are by no means complete, but are intended to provide an overview of the wide field of picosecond ultrasonics.

Picosecond ultrasonics signals are generally recorded in a pump-probe measurement configuration [1,66,67]. An ultrashort optical pump pulse creates a local stress/strain distribution in a sample caused by several possible excitation mechanisms [30]. This excitation launches an ultrasonic pulse or excites the acoustic eigenmodes of the structure. As an example, a propagating pulse then produces reflections at interfaces within the sample. By using time-delayed probe pulses in a pump-probe measurement setup, these returning strain waves can be observed via delay-dependent changes in reflectivity of the probe pulses. The time-of-flight of ultrasonic signals from the sample is mapped with the pump-probe measurement timing resolution, which is defined by the laser pulse duration [30].

The picosecond ultrasonics signals are tiny modulations on the reflected probe power generally ranging in $10^{-4}$-$10^{-6}$ level. Therefore, careful design of the experimental setup is required to achieve sufficient measurement precision. Originally, these measurements were implemented by using a single ultrashort pulse laser with a mechanical delay line and a high-frequency lock-in detector. High-frequency lock-in detection allows one to shift the data acquisition up to $f_{rep}/2$, where laser noise and other technical noise are generally low. Here $f_{rep}$ is the laser repetition rate. However, as measurements require timing range scans up to multiple nanoseconds, the delay line design becomes challenging [68].

An alternative approach is to use an equivalent time sampling (ETS). In ETS, the sampling is performed at a distinct rate to the excitation [69,70]. The specific implementation of such a configuration to pump-probe spectroscopy uses a pair of lasers with a locked repetition rate difference $\Delta f_{rep}$. This measurement configuration is also called asynchronous optical sampling (ASOPS) [71]. In this configuration, the pump-probe delay is swept over a range of $1/f_{rep}$ in a time given by $1/\Delta f_{rep}$. Consequently, the measurement time on the oscilloscope is related to the optical delay (the equivalent time) by the ratio $\Delta f_{rep}/f_{rep}$. Thus, ETS enables the transfer of optical time scales (femtosecond to nanosecond) to electronic time scales (nanosecond to millisecond), which can be easily recorded by an oscilloscope. The smallest sampling step $\Delta \tau$ is determined by $\Delta \tau = \Delta f_{rep}/f_{rep}^2$. For example, a pair of 80-MHz oscillators, with a slight repetition rate difference of 1 kHz, optical delay scans over 12.5-ns range will be obtained in 156-fs steps. This is equivalent to a mechanical delay stage moving at an ultra-high speed of 1.875 km/s over a 1.875-m distance. Significantly faster scanning speeds can be obtained with higher repetition rate lasers with the same timing resolution [72].

Typically, ETS (resp. ASOPS) can be implemented with a pair of modelocked lasers whose repetition rate difference is locked electronically [71]. Such laser systems are closely related to dual optical frequency combs (dual-combs) [73] in the sense that the delay increment between subsequent pairs of pulses is constant. An additional feature of dual-comb sources is that they exhibit a high degree of phase coherence between the two combs. However, in pump-probe measurements the sample response is typically proportional to pump intensity, so phase-coherence of the optical comb lines is not required.

A schematic illustration of how two different repetition rate pulse trains are used for picosecond ultrasonics measurements is shown in Fig. 1. The ETS technique was first applied to picosecond ultrasonics in 2006 [74], where significant measurement improvements compared to usual mechanical delay line-based pump-probe measurements were achieved. Since then, ETS with a locked laser pair became commonly used in the opto-acoustic community.

However, the key disadvantage in using a locked laser pair for ETS is that it requires two separate lasers, as well as piezoelectric actuators and locking electronics to stabilize the repetition rate difference. This dramatically increases the cost and complexity of the method and thus is difficult to adopt in a practical setting. Recently, we overcame this limitation by developing free-running dual-comb laser oscillators, which produce two pulse trains from the same laser cavity with an intrinsically stable and at the same time controllable repetition rate difference [75,76]. Since both combs are produced from the same laser cavity, they generally have significant phase coherence, often sufficient even for dual-comb spectroscopy [77,78] and

LIDAR [79] applications. In cases where phase and timing drifts are faster than the desired averaging time, adaptive sampling methods can usually be applied [80]. Dual-comb generation from free-running lasers has been widely explored in recent years, as reviewed in [81].

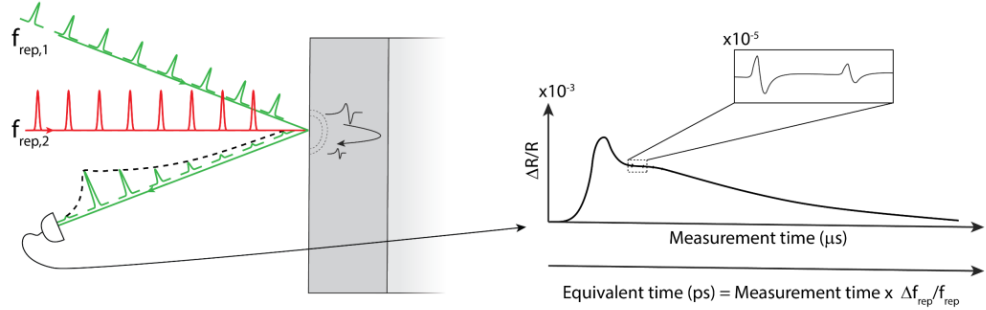

**Fig. 1.** Schematic illustration of a picosecond ultrasonics measurement with the pump and probe having two slightly different repetition rates $f_{rep,1}$ and $f_{rep,2}$. The pump (red) excites an ultrasonic wave in a sample. The returning reflection is then sampled with the probe (green). The signal is detected with a photodiode and recorded on an oscilloscope (right). Within the timescale of the thermal response of the sample, ultrasonic signatures can be observed (inset). The measurement time is related to the equivalent pump-probe time by the laser repetition rate difference and probe repetition rate ratio.

Lately, we have applied the polarization-multiplexing concept to powerful femtosecond solid-state lasers [82]. We have demonstrated a free-running dual-comb laser oscillator delivering 175-fs pulses at 137 MHz repetition rate with tunable 1-kHz repetition rate difference. The laser delivered up to 440 mW per comb and it was applied to ETS measurements of recovery dynamics in integrated semiconductor structures. Solid-state lasers are particularly interesting for rapid pump-probe measurements because they generally exhibit ultra-low noise at high sampling frequencies [83–86]. This performance is not matched by other commonly used laser technology platforms. Additionally, solid-state lasers can be configured for high repetition rates or high pulse energies at lower repetition rates [87,88].

Compared to our previous solid-state dual-comb demonstration [82], here we demonstrate a significantly performance-improved free-running dual-comb laser oscillator delivering 1.8 W of average power at 80-MHz repetition rate at a pulse duration of 110 fs from each output comb; this corresponds to a peak power enhancement of more than ten times. Although higher average power free-running dual-comb laser oscillators using thin-disk geometry have been demonstrated recently [89,90], here we demonstrate the highest power free-running dual-comb bulk oscillator to date. The bulk laser geometry is highly relevant for practical applications of such lasers as it enables compact and cost-effective optical sensor development.

We apply this laser to picosecond ultrasonics measurements and demonstrate shot-noise limited $1.8 \times 10^{-5}$ relative reflectivity change sensitivity in a single measurement trace without averaging. Since high sensitivity is important for this non-destructive testing technique, our results represent an important advance to the picosecond ultrasonics measurement method development and its application readiness.

In Section 2, we describe our 80-MHz dual-comb laser design and report details on its performance. In Section 3 we demonstrate the application of this laser for the measurement of superlattice phonons on a semiconductor multi-quantum-well (MQW) structure. In Section 4 we present metallic layer thickness and Brillouin scattering measurements. We compare these measurements to those obtained with a commercial locked laser pair. Finally, in Section 5 we elaborate on measurement system optimization steps towards optimal picosecond ultrasonics measurement performance.

## 2. Free-running 80-MHz dual-comb laser oscillator

In this section, we present technical design details of a high-power free-running 80-MHz dual-comb laser. In an end-pumped laser cavity configuration, the gain crystal is positioned next to and pumped through a flat output coupling mirror as shown in Fig. 2(a). We use a multimode pump diode which emits light centered at 980 nm and can deliver up to 20 W of average power (I5F1S15, Coherent Inc.). To pump both combs with this diode, we split the collimated output beam into two separate paths using a 50/50 beam splitter and combine them back on a D-shaped mirror with a slight horizontal position offset. This position is then re-imaged onto the gain crystal so that a collinear dual-focus is achieved. The diode beam was measured to have a $M^2$ value of 20 and it is focused onto the gain crystal to an estimated 64 μm beam radius ($1/e^2$ intensity).

Yb:$CaF_2$ is a well-suited gain crystal for polarization-multiplexed dual-comb lasers as it exhibits isotropic gain properties, which are desired for maximal noise correlation of the two combs. Furthermore, it has favorable thermal properties and exhibits a smooth and broadband gain spectrum [91]. However, since the laser crystal does not have a well-defined polarization axis, it is susceptible to thermally-induced depolarization effects which can degrade the oscillator performance. Fortunately, it was shown that if the crystal is [111] oriented, it can be aligned for minimal depolarization for two nearly orthogonal polarization planes [92]. For this experiment, we have grown a Yb:$CaF_2$ crystal with 4.5% at. Yb doping by using a Bridgman technique. A mixture of $CaF_2$, $YbF_3$ powders (purity: 4 N) was placed in a graphite crucible within the growth chamber. A good vacuum ($<10^{-5}$ mbar) was then realized, and the crystal growth was carried out with a pulling rate of 1–3 mm/h. After completing the growth process, the grown crystals were cooled down to room temperature within 72h. Finally, the crystal was polished to 4 mm length and antireflection coated for the pump and laser wavelengths. Using an external setup to characterize thermal stress-induced depolarization, similar to the one reported in [93], we have pre-selected the gain crystal so that it exhibits minimal thermal-stress induced birefringence and is appropriately pre-aligned to the polarization planes defined by the birefringent crystals.

The resonator is multiplexed to support two nearly-common path modes at two orthogonal polarizations by using a birefringent crystal multiplexing approach [75]. The cavity layout is shown in Fig. 2(a), and the output beam profiles are shown in Fig. 2(b). We use 6 mm-long calcite crystals, which are cut at 45° from the c-axis. Calcite was chosen deliberately for this laser oscillator as it exhibits a high spatial walk-off with minimal nonlinearity. Using one calcite crystal near the gain medium and one near the semiconductor saturable absorber mirror (SESAM) [94] leads to well-separated modes on the active elements of the cavity. We use wedged crystals to avoid etalon effects and arrange the wedge in such a way to increase the modal separation further and thus minimize any possible interaction between the beams. Finally, the two calcite crystals are oriented orthogonally to each other to minimize the roundtrip group delay difference experienced by the two polarization modes.

Modelocking of the oscillator is initiated and stabilized with a dielectric top-coated three quantum-well (QW) SESAM which has a high saturation fluence of 54 μJ/$cm^2$, $F_2$ parameter of 1.1 J/$cm^2$ and modulation depth of 1.6% (characterized at 1050 nm with 185-fs pulses). Self-starting and simultaneous mode-locking is achieved in a total pump power range from 6.8 W up to 11.2 W (Fig. 2(c)). The maximal output power where simultaneous modelocking is maintained is 1.8 W per comb, which corresponds to 32% optical-to-optical efficiency. Compared to the state-of-the-art Yb:$CaF_2$ bulk laser oscillator [95], our laser achieves similar output parameters and exhibits high efficiency despite having additional intracavity elements and being pumped by a low-brightness pump. Compared to our previously reported dual-comb laser oscillator [82], this new oscillator achieves significantly higher power and shorter pulses leading to peak power enhancement of more than 10. Such a high-peak-power laser is ideally suited for nonlinear spectroscopy and frequency conversion to other spectral domains. The laser output power is currently limited by self-phase modulation and the laser gain bandwidth. This limitation could be offset by using a higher output coupling rate, such that we reach the same minimum pulse duration at higher average power. The current output coupling rate of 5.5% was determined by the component availability. The output beam quality parameter $M^2$ was measured to be better than 1.05.

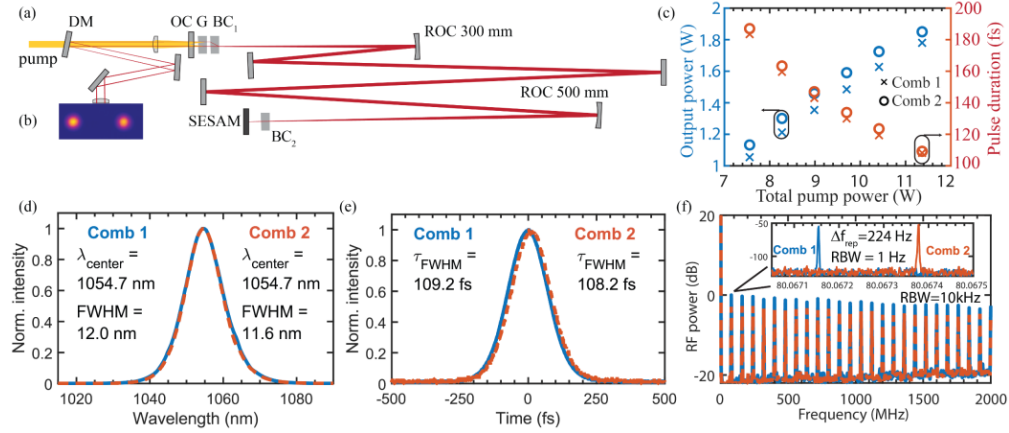

**Fig. 2.** (a) Layout of the 80-MHz dual-comb solid-state laser cavity. The gain medium (G) is a 4-mm long $CaF_2$ crystal with 4.5% at. Yb doping. The laser is pumped through the output coupler (OC). The two polarization states of the cavity are split by the insertion of two birefringent 45°-cut calcite crystals ($BC_1$ and $BC_2$) with 6-mm length, each. The laser output is separated from the pump beam with a dichroic mirror (DM) and then collimated by a lens. (b) Image of the output beams recorded with a WinCamD-LCM-NE 1" beam profiler. (c) Power slope and pulse duration evolution of both output beams when pumped simultaneously as a function of the total pump power (split with 1:1 ratio between the two pump spots). Panels (d) to (f) show the modelocking performance at 1.8 W output level per comb: (d) optical spectrum of each comb with center wavelength and full-width at half-maximum (FWHM) using a $sech^2$ fit. (e) pulse duration measurement using a second harmonic generation autocorrelator. The pulse duration FWHM is specified after deconvolution of the measured trace. (f) Radio-frequency (RF) spectrum of the laser. The inset shows high-spectral resolution measurements around the repetition rate of the laser. The 224-Hz repetition rate difference is indicated. RBW: resolution bandwidth.

Soliton modelocking [96] in the cavity is achieved by introducing a negative group delay dispersion of –3320 $fs^2$ per cavity round trip with dispersive mirrors. The output spectra (Fig. 2(d)) exhibit a hyperbolic secant shape with a high degree of spectral overlap due to the symmetric spectral gain profile experienced by both laser polarizations. The autocorrelation traces are depicted in Fig. 2(e). Figure 2(f) shows the radio-frequency spectra of the output pulse trains indicating clean fundamental modelocking. The inset shows a zoomed-in view of the fundamental harmonic where two distinct repetition rates can be resolved. In this measurement, the repetition difference was set to 224 Hz; however, it can be freely tuned by rotation of the second birefringent crystal.

For the ETS measurements discussed later in this publication, the timing resolution is limited by the detection bandwidth and not the delay step size. The resolution limit $\Delta\tau_D$ due to the available detection bandwidth ($BW$) is calculated by $\Delta\tau_D = \Delta f_{rep}/(f_{rep} \times BW)$. For example, in the case discussed later, an 11.6-MHz detection bandwidth and a 285-Hz repetition rate difference give the measurement resolution of 307 fs (Section 5). This is significantly more than the timing jitter we can expect from a free-running dual-comb laser. In fact, we have already shown that in polarization-multiplexed lasers, the period timing jitter (from trigger to trigger) can be less than 100 fs in a 10-second-long acquisition window [82] and that polarization-multiplexed dual-comb lasers can be used for phase-sensitive dual-comb spectroscopy experiments which require sub-cycle timing precision [77,78]. Precise and rapidly sampled relative timing jitter measurements go well beyond the scope of this publication, and further research is needed to develop an appropriate relative timing jitter sampling setup at a fixed repetition rate difference. However, we are confident that the timing jitter from the free-running dual-comb laser is not a limiting factor for picosecond ultrasonics measurements. Therefore, in the following we study slower drifts of the repetition rate difference.

The long-term fluctuations of the repetition rate difference can be measured relatively easily with a frequency counter. For this purpose, we use an optical cross-correlation setup to generate a difference frequency signal (more detail in the next section) and a frequency counter (Keysight, 53210A). We observe that the repetition rate difference is sensitive to the temperature of the birefringent crystals. The demonstrated laser is assembled on a breadboard without any isolating housing. Thus, the repetition rate difference stability was coupled to the laboratory conditions. Nevertheless, in stable laboratory conditions we obtain a highly stable repetition rate difference as shown in Figs. 3(a-b). The relative difference frequency stability $\sigma(\Delta f_{rep})/\Delta f_{rep}$ is $1.0 \cdot 10^{-5}$. We also calculate the Allan deviation and find that a minimum deviation in $\Delta f_{rep}$ of 0.41 mHz with respect to its mean is obtained for an 8-second averaging time window. In this case, the relative difference frequency stability is $2.2 \cdot 10^{-6}$. This therefore shows that the long-term relative timing between the two pulse trains remains well-defined and below the pulse duration.

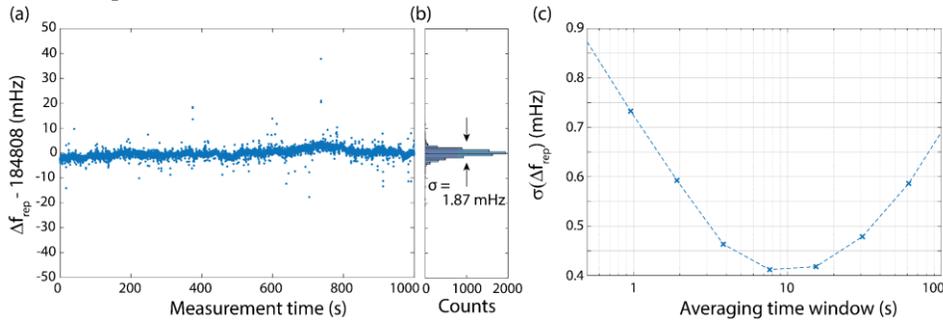

**Fig. 3.** (a) Measured repetition rate difference $\Delta f_{rep}$ stability between the two combs when sampled at ~8.3 Hz for 1000 seconds. (b) Repetition rate difference distribution on the data from (a). Calculated standard deviation was 1.87 mHz. (c) Allan deviation on the data from (a).

## 3. Acoustic waves in hybrid quantum well/superlattice structures

Picosecond ultrasonics has attracted significant attention in studying coherent acoustic phonons in epitaxially grown structures. For example, a direct application of such studies is the determination of the structural parameters in quantum cascade lasers [17]. The study of these phonons, which can reach THz frequencies [51,97–101], impacts the fields of opto-acoustics and mechanics [58,102–104], potentially yielding insights into advanced thermoelectric materials [105].

Recently, our work also covered hybrid structures that consist of QW regions and an underlying superlattice [106,107]. To demonstrate that a free-running dual-comb laser is well-suited for such studies, we prepare an experimental pump-probe setup as shown in Fig. 4(a). Cross-correlation between the two combs in a sum-frequency generation (SFG) crystal is used to generate a trigger signal. The SFG signal is detected with an amplified photodiode (PDA55, Thorlabs Inc.). This trigger setup is used in all further picosecond ultrasonics measurements presented in this paper, which involve a free-running dual-comb laser.

We use half-wave plate and polarizer pairs to control the pump and probe powers. We first perform a picosecond ultrasonics experiment in a non-collinear interaction geometry (Fig. 4(a)). The pump and probe are focused on the sample with a 20-mm focal length aspheric lens. We obtain 8.5-μm pump and probe beam radii ($1/e^2$) at the sample position. Since the beams arrive onto the sample non-collinearly, the reflected pump and probe beams are spatially separated. We use a D-shaped pick-up mirror to collect the probe light and send it on a photodiode. For the following experiment, we have used an amplified photodiode (PDA10CF Thorlabs Inc.).

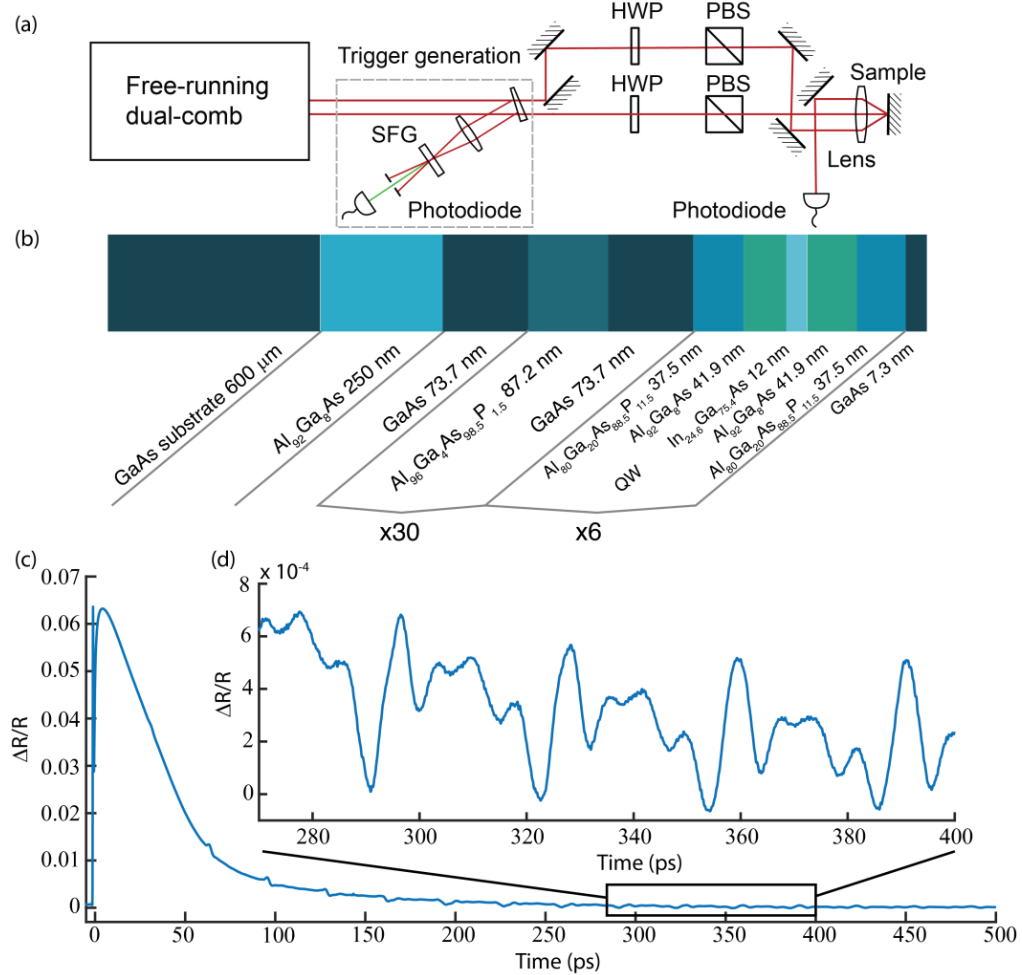

Fig. 4. (a) Equivalent time sampling (ETS) setup using a free-running dual-comb laser. SFG – sum-frequency generation crystal; HWP – half-waveplate, PBS – polarizing beam splitter. (b) Semiconductor sample: multi-quantum-well SESAM structure. (c) ETS measurement on SESAM response in case of 3.15 mJ/cm$^2$ average fluence excitation. Here $10^4$ trace averages were accumulated. (d) Zoom-in of the ultrasonic response.

We perform pump-probe measurements on a SESAM sample with a six QW absorber. The exact SESAM structure is shown in Fig. 4(b). The sample was excited with 572 mW of pump power corresponding to 3.15 mJ/cm$^2$ average fluence excitation. This fluence is close to, but below, the damage threshold. The reflected probe power reaching the photodiode was 0.31 mW. The repetition rate difference was set to 255.54 Hz. The photodiode signal was filtered with an analog 30-MHz low-pass filter and was recorded on the oscilloscope (WavePro 254HD, Teledyne LeCroy) at 100 MS/s sampling rate with a 20-MHz analog bandwidth setting. Using the oscilloscope and the trigger setup, we perform $10^4$ trace averages. The first 500 ps of the resulting signal after a 10-MHz low-pass digital filter are shown in Fig. 4(c). The time-domain signal shows a fast initial SESAM response followed by a relatively slow carrier recombination [108] with superimposed contributions caused by the excited acoustic wave dynamics in the sample, highlighted in the zoom-in of Fig. 4(d).

Previous results of SESAM samples have revealed the complex acoustic wave excitation, propagation, and detection mechanisms in these structures [107,109,110]. Here, we will focus on the extraction of the structural and material properties to demonstrate the capabilities of the new measurement system.

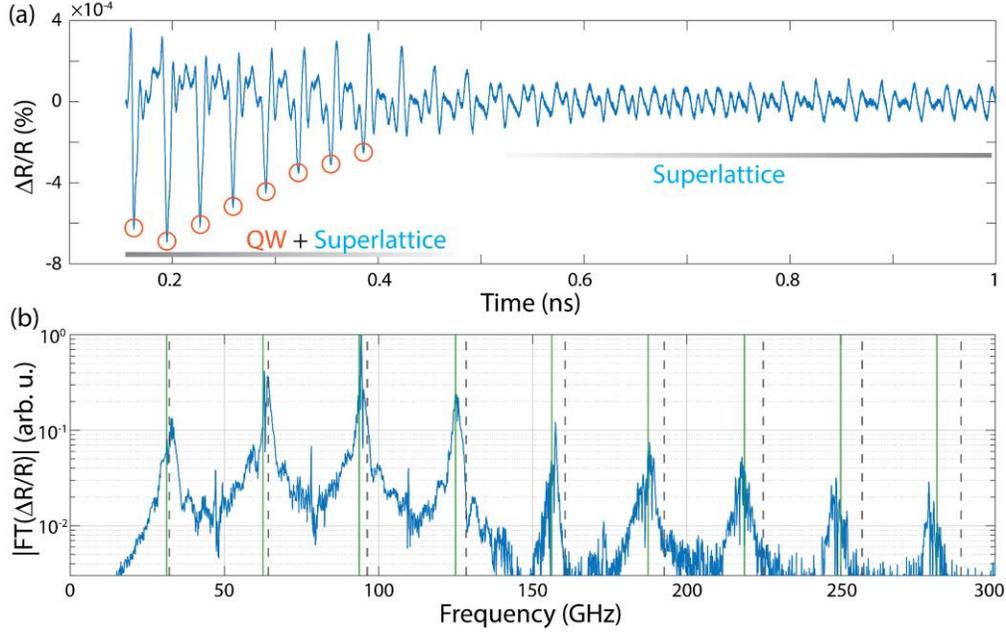

**Fig. 5.** (a) Extracted acoustic contribution to the time domain signal showing the regions where the signal from the QW and superlattice as well as the superlattice alone are present. Orange circles mark the pulses used for the determination of the pulse propagation times. (b) Fourier transformation of the time domain data up to about 5 ns as solid blue line. Gray dashed lines indicate the position of Brillouin zone center (BZC) modes of the superlattice for nominal values in thickness and sound velocities using AlGaAs sound velocity for the AlGaAsP layers. Solid green lines indicate the adjusted dispersion with 5.3% smaller sound velocities of the AlGaAsP layers.

**Table 1. Sound velocities of materials in the SESAM sample.**

|  | GaAs | InAs | $In_{24.6}Ga_{75.4}As$ | $Al_{96}Ga_4As$ | $Al_{80}Ga_{20}As_{88.5}P_{11.5}$ (this work) |
|---|---|---|---|---|---|
| Longitudinal sound velocity in (100) direction (m/s) | 4730 (ref. [111]) | 3830 (ref. [112]) | 4509 (interpolation) | 5640 (ref. [111]) | 5250 (Pulse) 5308 (Superlattice) |

Since the sample consists of a QW region and an underlying superlattice structure, characteristic signatures from both regions are expected to be in the time domain signal. After removal of the slowly varying background by a moving average we obtain the acoustic signal shown in Fig. 5(a). Two regions can be identified: the first 500 ps show a combination of acoustic pulses and oscillatory components. The latter, region 2, persist beyond the shown time window up to about 5 ns. We will first turn to the discussion of the acoustic pulses. These originate from the QW region of the sample [107,109], which is corroborated by the time separation of subsequent pulses at $32.0 \pm 0.2$ ps. This is in very good agreement with the calculated propagation time from the middle of one QW to the middle of another nearby QW being 31.3 ps neglecting possible changes in the sound velocity of AlGaAsP, i.e. calculated for AlGaAs. The sound velocities used for the calculations are listed in Table 1.

If we take changes in the sound velocity of the AlGaAsP into account and adjust the sound velocity to the experimental propagation times, we obtain the sound velocity in the AlGaAsP layer which is also given in Table 1. marked with (Pulse). Unfortunately, there is a lack of literature data for the AlGaAsP compound which prevents a comparison.

Similar to previous work [107,109], the acoustic pulses are thus mainly generated and detected in the QW's. This is a good demonstration of the setup capabilities to investigate QW-based structures.

We will now discuss the oscillatory components present in the signal. A Fourier transformation of the time data is shown as a blue solid line in Fig. 5(b). A regular frequency pattern emerges up to around 300 GHz that exhibits typical superlattice characteristics. Determination of the acoustic response of the superlattice structures often requires a detailed calculation of the structure's acoustic dispersion. In such calculations, special points of interest are the acoustic bandgap openings at the Brillouin zone center (BZC) and Brillouin zone edge (BZE). However, our superlattice exhibits a large two-layer period thickness of 160.9 nm. This results in small bandgaps [37], with a maximum frequency gap value of 1.6 GHz, where the additional phosphorus content of the AlGaAs layer was neglected for this estimate. Due to the small bandgaps, the acoustic dispersion of the superlattice can be linearized similar to the work by Ezzahri et al. [37]. This is equivalent to the calculation of the mid-bandgap frequencies at either BZC or BZE. Taking into account the zone folding introduced by the periodicity of the superlattice structure, the BZC and BZE modes are then located at frequencies $f$ given by:

$$f(n) = n \frac{v_{eff}}{2 d_{sl}},$$

where $n$ is an integer. BZC modes are calculated for even $n$ and BZE modes for uneven $n$. The effective sound velocity $v_{eff}$ is given by the weighted average of the layers:

$$\frac{d_{sl}}{v_{eff}} = \frac{d_A}{v_A} + \frac{d_B}{v_B}.$$

Here $d_{sl} = d_A + d_B$ gives the superlattice period thickness while A and B denote the GaAs and AlGaAsP layers respectively.

Taking the respective concentrations of the compounds into account and using literature values for the sound velocities, we obtain BZC modes at frequencies indicated by the dashed grey lines in Fig. 5(b). Here again, we used AlGaAs for the calculations.

Several higher-order modes are visible in the spectrum and a clear mismatch becomes apparent at higher frequencies. Since the layer thickness was verified by x-ray diffraction measurements the most likely source of this mismatch stems from the uncertainty in the sound velocity of the AlGaAsP layers.

After extracting the mode frequencies and performing a least-squares fit for the AlGaAsP sound velocity, we obtain the solid green lines as the final result. These show a much better agreement with the experimentally obtained mode frequencies. The resulting sound velocity is given in Table 1, marked with (Superlattice). The results for the sound velocity determined from the QW and the superlattice region only deviate by 2.5% from each other despite being determined independently.

This demonstrates that the new measurement system performs well in the investigation of acoustic superlattice structures and accompanying high-frequency acoustic modes.

## 4. Acoustic echoes and time-resolved Brillouin scattering

In this section, we present a series of picosecond ultrasonics experiments where broadband acoustic echoes and time-resolved Brillouin scattering signals are recorded from different interaction geometries in the same sample with two different measurement systems. The first set of experiments was performed using the free-running dual-comb laser, which was presented in Section 2. The second set of experiments was performed using a commercial laser system containing a pair of modelocked lasers with an electronically locked repetition rate difference (Amplitude Systems). The commercial system delivers two pulse trains with a 330-fs pulse duration at a 48-MHz repetition rate and a center wavelength of 1030 nm [26]. The repetition rate difference for this system was set to 994 Hz. The beam used as the probe was frequency doubled to 515 nm.

For all experiments, we used the same 1-mm-thick $SiO_2$ sample, which was metalized with a tungsten (W) layer. Depending on the experiment, the pump and probe light interacted with the sample either from the air-W interface or through the air-$SiO_2$ interface. In the experiments, we have used the same focusing microscope objective with 10x magnification and a numerical aperture of 0.25. The objective was anti-reflection coated for green light. The experimental setup used with the free-running dual-comb laser is laid out in Fig. 6(a). A conceptually equivalent setup was used for the experiments with the commercial pair of stabilized laser system.

### 4.1 Acoustic echoes

Acoustic echo signals are recorded using the sample configuration shown in Fig. 6(b). The pump pulse (1030 nm or 1055 nm) is used to launch coherent phonons in the tungsten layer via optical absorption. This creates a broadband elastic strain pulse (a wavepacket of longitudinal acoustic phonons) which then travels through the sample and is reflected by the W-$SiO_2$ interface. The returning echo then modulates the density of the metallic surface, which creates a refractive index modulation. This leads to a modulation in reflectivity which is sampled by the delayed probe pulses (515 nm or 527 nm).

To record these acoustic echoes and infer the tungsten thin-film thickness, we used the previously mentioned experimental setup (Fig. 6(a)). First, we describe the measurement taken with the free-running dual-comb laser. For this experiment, we have used 49.6 mW of pump power (1055 nm) focused to an estimated average fluence of 373 μJ/cm$^2$. The probe power (527 nm) was adjusted so that 0.3 mW was incident on a balanced detector (PDB425A, Thorlabs Inc.). We used an additional copy of the probe beam sampled before the interaction with the target to balance the optical powers on the detector and thus to obtain minimal DC voltage on the output. By using the probe power measurement on the detector and the tabulated detector responsivity with transimpedance gain, the DC level for the case of blocked balancing light can be estimated. The measured AC signal is then divided by the estimated DC level and the relative reflectivity change ΔR/R quantity is obtained. For this measurement, we have set the laser repetition rate difference to 1021.69 Hz. The balanced detector output signal was recorded using a data acquisition card (PCI-5122, National Instruments) with a 100 MS/s sampling rate. The analog signal was filtered with a 30-MHz low-pass filter. Using the trigger setup described above, we collected $10^5$ trace averages. Figure 6(c) (blue line) shows the recorded trace using the free-running dual-comb laser. In this measurement, a digital 5-MHz low-pass frequency filter was applied.

We performed a similar measurement in another laboratory on the same sample using the commercial system described above. In this measurement we have used a slower (PDB440A, Thorlabs Inc.) balanced amplified photodetector which limits the detection bandwidth to 15 MHz. The full aperture of the same microscope objective was illuminated with the pump and probe beams. The pump power on the sample was measured to be 65 mW and the probe power on the detector was 1.0 mW. Using a data acquisition card (PXI-5122, National Instruments) with a 33-MHz sampling rate triggered data collection and averaging of $1.1 \times 10^5$ traces was performed. The final recorded trace is shown in Fig. 6(c) (red line). In this measurement set, the trigger signal was generated from a cross-correlation signal of both pulse trains generated by two-photon absorption on a GaAsP photodiode [26].

Both measurements (with a free-running dual-comb and the locked laser pair) are shown on the same time axis for direct comparison. From this, we can see that both measurement techniques allowed us to record nearly identical picosecond ultrasonics signals on top of a thermal background [113]. In the measurement, we can see up to three acoustic echoes. The delay between the first and the second echo is 101.6 ps extracted from the local minima in the signal. With the longitudinal acoustic wave velocity in tungsten of 5.2 nm/ps [114], the thin-film thickness can be estimated to be 264 nm. This agrees well between the two measurements. The measurement accuracy is not limited by the timing jitter of the free-running dual-comb laser since the signal is much slower than the measurement resolution. A more detailed signal analysis which goes beyond the scope of the paper is needed to obtain an accurate measurement uncertainty estimate.

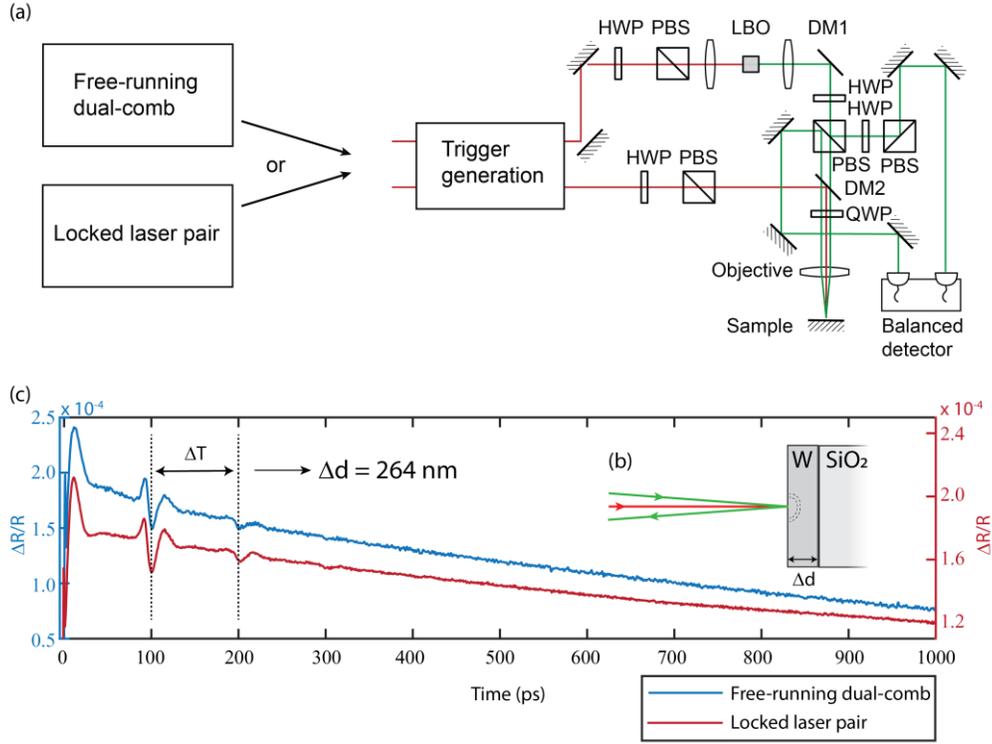

**Fig. 6.** (a) Experimental setup used for picosecond ultrasonics measurements on a metalized sample with the free-running dual-comb laser. HWP – half-waveplate, PBS – polarizing beam splitter, DM1 – dichroic mirror: highly reflective (HR) for 527 nm and highly transmissive (HT) for 1055 nm. DM2 - dichroic mirror: HR 1055 nm, HT 527 nm. QWP – quarter waveplate (for 527 nm), LBO – lithium triborate used for second harmonic generation. Balanced detector: PDB425A (Thorlabs Inc.). An equivalent setup was used for measurements with a locked laser pair and the key difference was that another balanced detector PDB440A (Thorlabs Inc.) was used. (b) Interaction with the sample configuration. Green and red illustrate probe and pump beams, respectively. (c) Recorded relative reflectivity change using two different experimental setups. The blue line (left y-axis) indicates the measurement result obtained with a free-running dual-comb laser and the red line (right y-axis) indicates the measurement result obtained with the locked laser pair.

*4.2 Time-resolved Brillouin scattering*
Brillouin scattering is an interaction between light and phonons [115]. The picosecond ultrasonic technique allows the detection in the time domain of the Brillouin interaction with coherent phonons [29]. We study time-resolved Brillouin scattering experimentally by using the interaction geometry shown in Fig. 7(a). The pump (1030 nm or 1055 nm) excites acoustic waves in the $SiO_2$ by absorption at the $SiO_2$-W interface. Then the incoming probe beam interacts with the acoustic waves in the bulk $SiO_2$, which causes scattering in the backward direction. The scattering is most efficient if a phase-matching condition is satisfied: $\boldsymbol{k_i} + \boldsymbol{q} = \boldsymbol{k_s}$.

Here $k_i$ is the incident photon momentum, $k_s$ is the scattered photon momentum, and $q$ is the acoustic phonon momentum (vector quantities). Since $|k_i|$ and $|k_s|$ are much larger than $|-q|$, the scattering in the backward direction is given approximately by $k_s = -k_i$. This relation leads to a phase-matching solution for acoustic phonons with a Brillouin frequency $f_B = 2v_{ac}n/\lambda$, where $v_{ac}$ is the acoustic wave velocity, $n$ is the refractive index of the medium and $\lambda$ is the probe wavelength. Therefore, a measurement of $f_B$ allows the acoustic wave velocity to be determined independently of the dimensions of the sample.

Similar to Section 4.1, we perform a time-resolved Brillouin scattering experiment using the setup shown in Fig. 6(a). In the case of the free-running dual-comb laser-driven experiment, we have used 30.3 mW of 1055 nm pump power focused to an average fluence of 226 µJ/cm². The repetition rate difference was set at 285.52 Hz. As opposed to the measurement shown in Section 4.1, which utilized a balanced photodetector, in this measurement we use a single photodiode (the exact data acquisition scheme is motivated and explained in Section 5). The 527 nm probe power incident on the photodiode was 3 mW. The signal was recorded with the oscilloscope (WavePro 254HD, Teledyne LeCroy) at 100 MS/s sampling rate with 20-MHz analog bandwidth setting. 5172 traces were averaged to obtain the time-resolved Brillouin scattering measurement shown in Fig. 7(b). For this measurement, a digital 2-MHz low-pass filter was applied.

Analogous to Section 4.1, we have performed an independent measurement in another laboratory on the same sample using the commercial pair of laser system discussed above. In this measurement, as before, we have used a (PDB440A, Thorlabs Inc.) balanced amplified photodetector. The pump power on the sample was measured to be 65 mW and the probe power reaching the detector was 2.0 mW. Averaging of $1.1\times10^5$ traces was performed. The final recorded trace is shown in Fig. 7(c).

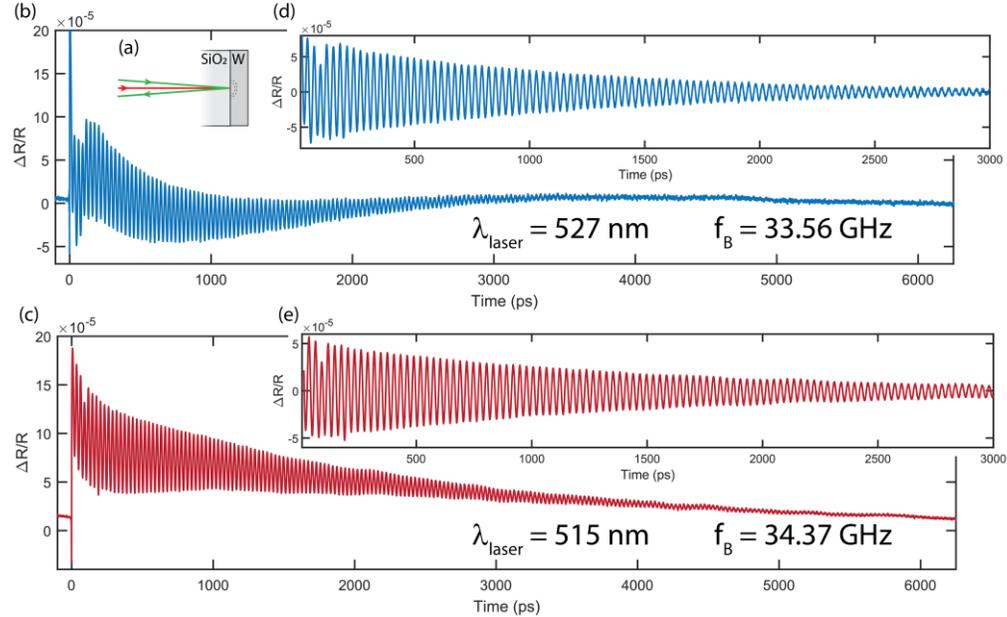

**Fig. 7.** (a) Interaction geometry used in the experiment. (b) Time-resolved Brillouin scattering recorded with the free-running dual-comb laser. For this measurement, the detection technique described in Section 5 was used and any pump-independent probe voltage offset was subtracted. Here 5172 traces were accumulated in 20-second-long acquisition. (c) Time-resolved Brillouin scattering recorded using the locked laser pair. For this measurement, a balanced photodetector (PDB440A, Thorlabs Inc.) was used. Here $1.1\times10^5$ traces were accumulated. Insets (d) and (e) show the recorded signals with the slow-moving thermal background subtracted.

When comparing the two measurements, it appears that the slow-varying thermal parts of the signals are quite different. This is due to the detection method used in the free-running dual-comb laser-based measurement, which is explained in Section 5. We have used a bias-T to separate the AC and DC parts of the signal, and this separation has removed some low-frequency components. Hence, the slow-varying signal shape was slightly altered. Nevertheless, by applying a moving average filter we can extract the rapid oscillation signals free from the thermal decay, as shown in Fig. 7(d) and 7(e). In this case, the traces look very similar in amplitude and time. By applying a Hann time-window and performing Fourier transformation, we find Brillouin oscillation frequencies of 33.56 GHz and 34.37 GHz, respectively for the two measurements. If the probe wavelength difference is considered, the measured frequencies are different only by 0.08%. This therefore demonstrates that a free-running dual-comb laser can be used for picosecond ultrasonics measurements, as we have quantitatively obtained the same results as measured with a commercial ASOPS system containing a pair of modelocked lasers locked to slightly different repetition rates.

## 5. Noise considerations

In this section we discuss the picosecond ultrasonics measurement noise optimization aspects and how they relate to the laser performance. First, we attempt to determine the measurement noise floor using our free-running dual-comb laser. As discussed in the introduction, solid-state lasers generally have ultra-low relative intensity noise (RIN) at high measurement frequencies. This advantageous property makes such lasers well-suited for ETS measurements. Therefore, the ETS measurements are likely to be shot-noise limited if a solid-state laser is used as previously demonstrated [116]. Based on this argument, balanced photodetectors are not necessary for picosecond ultrasonics measurements if a sufficiently low-noise laser is used because they do not suppress shot noise.

We therefore optimize our measurement system for the increased power on the detector to lower the shot noise contribution. We use a reverse-biased 0.8-mm$^2$ InGaAs photodiode, and observe that the photodiode response remains linear to pulses even with average powers above 15 mW in the case of 50-Ω termination. We then separate the DC and AC parts of the signal using a bias-T (BT45R, SHF Communication Technologies AG). The DC part is terminated with a 50-Ω resistor, and the voltage over this resistor is monitored with a voltmeter. The AC part of the signal is first filtered with an analog 30-MHz low-pass filter and then it is amplified with a variable-gain ultra-wideband voltage amplifier (DUPVA-1-70, Femto Messtechnik GmbH). The amplified signal is filtered again with a 30-MHz low-pass filter and finally recorded on a 50-Ω terminated oscilloscope. On the oscilloscope we use an analog bandwidth of 20 MHz and a sampling rate of 100 MS/s. The voltage gain in the detection electronics is calibrated by a sinusoidal 1-MHz signal from a frequency generator. This allows us to relate the measured DC voltage to the AC part of the signal such that *ΔR/R = AC / (DC × Gain)*. The amplification gain is chosen such that no signal saturation occurs.

We use this optimized detection scheme to determine the measurement noise floor by detecting the probe (1055 nm) light incident on the photodiode (no pump on the sample). In this case we have used 70 dB power amplification gain. We then acquire a 20-second-long trace which we average based on the timing of the trigger signal. The noise floor of the measurement is calculated from the standard deviation of the recorded ΔR/R signal. The inverse of the signal-to-noise ratio (SNR) is shown in Fig. 8 as a function of the number of averages. Each average corresponds to a measurement time of *1/Δf*$_{rep}$.

In, addition, we also calculate noise-floor from the shot-noise contribution. The shot-noise variance on a photodiode signal is determined by a well-known formula $\sigma^2 = 2qI_{avg}BW$, where $q$, $I_{avg}$, $BW$ are the elementary charge, average photocurrent and the detection bandwidth, respectively. The photocurrent $I_{avg}$ is generated by average optical power $P_{avg}$ on the detector with responsivity $R(\lambda)$. The relative shot-noise contribution on the signal then is calculated by taking a ratio $\sigma/I_{avg}$. Hence, the shot-noise limit on the measurement is given by the following equation:

$$1/SNR_{shot-noise} = \sqrt{\frac{2qBW}{P_{avg}R(\lambda)}}$$

We find that our free-running dual-comb ETS measurements are shot-noise limited even at 15.3 mW of probe power on the detector as can be seen in Fig. 8. At these conditions, we obtain the measurement noise floor of $1.8\times10^{-5}$ for the detection bandwidth of 11.6 MHz. This corresponds to -165 dBc/Hz RIN at such high frequencies. We note that this is still shot-noise limited due to the photodiode limitations and the actual laser noise floor is not yet determined.

In this picosecond ultrasonics demonstration, we have used an 80-MHz repetition-rate dual-comb laser. Thus, in a single measurement a 12.5 ns time window gets sampled. If the target response is significantly shorter, not all acquisition time is useful. Further measurement speed improvements can be obtained by using a higher repetition rate laser. For instance, if only a 4 ns time window is relevant, then a 250-MHz repetition rate laser can be used and the data acquisition rate can be further increased, for instance to 10 kHz if a 160-fs timing step is required.

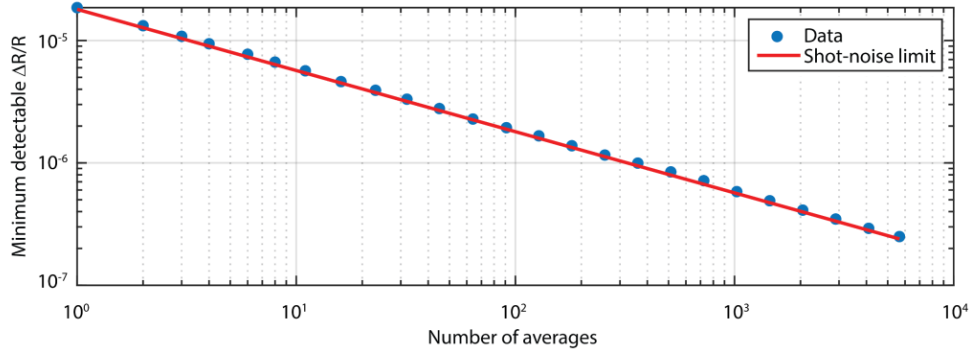

**Fig. 8**. The achievable signal-to-noise ratio (SNR) measurement with respect to the number of averaged traces. The standard deviation was calculated for a 100-μs long interval per averaged trace. The measured standard deviation on the data is close to the calculated shot-noise limit (red line). Therefore, we conclude that the minimum detectable signal is determined by the shot-noise. In this measurement, the detection bandwidth was set to 11.6 MHz and average power on the detector was 15.3 mW (1055 nm). We obtain $2.5\times10^{-7}$ sensitivity with 5659 traces in a 20-second-long acquisition.

## 6. Conclusions

Picosecond ultrasonics is a powerful non-destructive testing technique which allows one to probe thin-films with sub-nm precision. However, to this date, the practical adoption of this method was challenging due to the requirements for long mechanical delay stages and high-frequency lock-in detectors or by the cost and complexity of commercially available systems using two pairs of actively stabilized modelocked lasers. In this paper, we have shown a novel free-running dual-comb laser which delivers two combs with more than 1.8 W of average power and 110-fs pulse duration at an 80-MHz repetition rate from a single laser cavity. The two combs have slightly different repetition rates such that high-precision ETS measurements can be performed. The demonstrated laser has more than ten times greater peak power compared to our previously reported free-running dual-comb laser oscillator [82] and is the highest average power solid-state bulk free-running dual-comb oscillator to date.

We have applied this novel laser for picosecond ultrasonics measurements in a pump-probe measurement configuration. First, we have demonstrated measurements in a non-collinear configuration on an MQW-SESAM where superlattice phonons were observed. The results were compared to data acquired via x-ray diffraction measurements allowing to extract speed of sound in AlGaAsP. Furthermore, we have also demonstrated picosecond ultrasonics measurement on a metalized sample in a two-color collinear experiment. We have observed broadband acoustic echoes and Brillouin scattering signals. These measurements were compared to results acquired with a much more complex commercial system with two stabilized modelocked lasers showing a high-degree of agreement. Finally, due to the ultra-low RIN noise of our solid-state laser at high sampling frequencies, we were able to replace a typically-used balanced detector in the measurement setup with a single photodiode configured for high-power recording. We also showed that shot-noise limited data acquisition can be performed with a $1.8\times10^{-5}$ relative reflectivity change sensitivity per single trace in the case of a 307-fs sampling resolution. This effectively means that a 1-ns time window is scanned with this high sensitivity in a time of only 280 μs. Therefore, we conclude that a free-running solid-state dual-comb laser is an ideally-suited tool for picosecond ultrasonics and other similar pump-probe measurements.

**Funding.** This work was supported by a BRIDGE Discovery Project Nr. 40B2-0_180933 a joint research programme of the Swiss National Science Foundation (SNSF) and Innosuisse – the Swiss Innovation Agency.

**Acknowledgments.** We thank FIRST clean room facility at ETH Zurich where SESAM samples were prepared. F. B., and B. A. acknowledge financial support from the Agence Nationale de la Recherche (grant ANR-17-CE11-0020-01).

**Disclosures.** The authors declare no conflicts of interest.

**Data availability.** Data underlying the results presented in this paper is available at ETH Zurich Research Collection library.

## References

1. C. Thomsen, J. Strait, Z. Vardeny, H. J. Maris, J. Tauc, and J. J. Hauser, "Coherent Phonon Generation and Detection by Picosecond Light Pulses," Phys. Rev. Lett. **53**(10), 989–992 (1984).
2. L. Rouaï, B. Bonello, G. Louis, B. Perrin, and P. Peretti, "Elasticity of ultrathin copper-phthalocyanine Langmuir–Blodgett films by picosecond ultrasonics," J. Appl. Phys. **85**(12), 8155–8159 (1999).
3. H. Ogi, N. Nakamura, and M. Hirao, "Picosecond ultrasound spectroscopy for studying elastic modulus of thin films: A review," Nondestruct. Test. Eval. **26**(3–4), 267–280 (2011).
4. C. Rossignol, B. Perrin, B. Bonello, P. Djemia, P. Moch, and H. Hurdequint, "Elastic properties of ultrathin permalloy/alumina multilayer films using picosecond ultrasonics and Brillouin light scattering," Phys. Rev. B - Condens. Matter Mater. Phys. **70**(9), 1–12 (2004).
5. B. Bonello, B. Perrin, E. Romatet, and J. C. Jeannet, "Application of the picosecond ultrasonic technique to the study of elastic and time-resolved thermal properties of materials," Ultrasonics **35**(3), 223–231 (1997).
6. A. Nagakubo, H. T. Lee, H. Ogi, T. Moriyama, and T. Ono, "Elastic constants of beta tungsten thin films studied by picosecond ultrasonics and density functional theory," Appl. Phys. Lett. **116**(2), 021901 (2020).
7. F. Xu, L. Belliard, D. Fournier, E. Charron, J.-Y. Duquesne, S. Martin, C. Secouard, and B. Perrin, "Complete elastic characterization of lithium phosphorous oxynitride films using picosecond ultrasonics," Thin Solid Films **548**, 366–370 (2013).
8. C. Mechri, P. Ruello, J. M. Breteau, M. R. Baklanov, P. Verdonck, and V. Gusev, "Depth-profiling of elastic inhomogeneities in transparent nanoporous low-k materials by picosecond ultrasonic interferometry," Appl. Phys. Lett. **95**(9), 091907 (2009).
9. H. D. Boggiano, R. Berté, A. F. Scarpettini, E. Cortés, S. A. Maier, and A. V. Bragas, "Determination of Nanoscale Mechanical Properties of Polymers via Plasmonic Nanoantennas," ACS Photonics **7**(6), 1403–1409 (2020).
10. J. Cuffe, O. Ristow, E. Chávez, A. Shchepetov, P.-O. Chapuis, F. Alzina, M. Hettich, M. Prunnila, J. Ahopelto, T. Dekorsy, and C. M. Sotomayor Torres, "Lifetimes of Confined Acoustic Phonons in Ultrathin Silicon Membranes," Phys. Rev. Lett. **110**(9), 095503 (2013).
11. C. J. Morath, G. Tas, T. C. D. Zhu, and H. J. Maris, "Phonon attenuation in glasses studied by picosecond ultrasonics," Phys. B Condens. Matter **219**–**220**(1–4), 296–298 (1996).
12. B. C. Daly, K. Kang, Y. Wang, and D. G. Cahill, "Picosecond ultrasonic measurements of attenuation of longitudinal acoustic phonons in silicon," Phys. Rev. B - Condens. Matter Mater. Phys. **80**(17), 1–5 (2009).
13. H. N. Lin, R. J. Stoner, H. J. Maris, and J. Tauc, "Phonon attenuation and velocity measurements in transparent materials by picosecond acoustic interferometry," J. Appl. Phys. **69**(7), 3816–3822 (1991).
14. T. C. Zhu, H. J. Maris, and J. Tauc, "Attenuation of longitudinal-acoustic phonons in amorphous SiO2 at frequencies up to 440 GHz," Phys. Rev. B **44**(9), 4281–4289 (1991).
15. J.-Y. Duquesne and B. Perrin, "Ultrasonic attenuation in a quasicrystal studied by picosecond acoustics as a

function of temperature and frequency," Phys. Rev. B **68**(13), 134205 (2003).
16. D. Brick, E. Emre, M. Grossmann, T. Dekorsy, and M. Hettich, "Picosecond Photoacoustic Metrology of SiO2 and LiNbO3 Layer Systems Used for High Frequency Surface-Acoustic-Wave Filters," Appl. Sci. **7**(8), 822 (2017).
17. A. Bruchhausen, J. Lloyd-Hughes, M. Hettich, R. Gebs, M. Grossmann, O. Ristow, A. Bartels, M. Fischer, M. Beck, G. Scalari, J. Faist, A. Rudra, P. Gallo, E. Kapon, and T. Dekorsy, "Investigation of coherent acoustic phonons in terahertz quantum cascade laser structures using femtosecond pump-probe spectroscopy," J. Appl. Phys. **112**(3), 033517 (2012).
18. M. Schubert, H. Schaefer, J. Mayer, A. Laptev, M. Hettich, M. Merklein, C. He, C. Rummel, O. Ristow, M. Großmann, Y. Luo, V. Gusev, K. Samwer, M. Fonin, T. Dekorsy, and J. Demsar, "Collective Modes and Structural Modulation in Ni-Mn-Ga(Co) Martensite Thin Films Probed by Femtosecond Spectroscopy and Scanning Tunneling Microscopy," Phys. Rev. Lett. **115**(7), 076402 (2015).
19. R. Côte and A. Devos, "Refractive index, sound velocity and thickness of thin transparent films from multiple angles picosecond ultrasonics," Rev. Sci. Instrum. **76**(5), 053906 (2005).
20. T. Pezeril, N. Chigarev, D. Mounier, S. Gougeon, P. Ruello, J.-M. Breteau, P. Picart, and V. Gusev, "Lumped oscillations of a nanofilm at adhesion bond," Eur. Phys. J. Spec. Top. **153**(1), 207–210 (2008).
21. M. Grossmann, M. Schubert, C. He, D. Brick, E. Scheer, M. Hettich, V. Gusev, and T. Dekorsy, "Characterization of thin-film adhesion and phonon lifetimes in Al/Si membranes by picosecond ultrasonics," New J. Phys. **19**(5), 053019 (2017).
22. M. Hettich, A. Bruchhausen, S. Riedel, T. Geldhauser, S. Verleger, D. Issenmann, O. Ristow, R. Chauhan, J. Dual, A. Erbe, E. Scheer, P. Leiderer, and T. Dekorsy, "Modification of vibrational damping times in thin gold films by self-assembled molecular layers," Appl. Phys. Lett. **98**(26), 261908 (2011).
23. J. D. G. Greener, E. De Lima Savi, A. V. Akimov, S. Raetz, Z. Kudrynskyi, Z. D. Kovalyuk, N. Chigarev, A. Kent, A. Patané, and V. Gusev, "High-Frequency Elastic Coupling at the Interface of van der Waals Nanolayers Imaged by Picosecond Ultrasonics," ACS Nano **13**(10), 11530–11537 (2019).
24. G. Tas, R. J. Stoner, H. J. Maris, G. W. Rubloff, G. S. Oehrlein, and J. M. Halbout, "Noninvasive picosecond ultrasonic detection of ultrathin interfacial layers: CF x at the Al/Si interface," Appl. Phys. Lett. **61**(15), 1787–1789 (1992).
25. M. Hettich, K. Jacob, O. Ristow, C. He, J. Mayer, M. Schubert, V. Gusev, A. Bruchhausen, and T. Dekorsy, "Imaging of a patterned and buried molecular layer by coherent acoustic phonon spectroscopy," Appl. Phys. Lett. **101**(19), 191606 (2012).
26. A. Abbas, Y. Guillet, J.-M. Rampnoux, P. Rigail, E. Mottay, B. Audoin, and S. Dilhaire, "Picosecond time resolved opto-acoustic imaging with 48 MHz frequency resolution," Opt. Express **22**(7), 7831 (2014).
27. S. Ramanathan and D. G. Cahill, "High-resolution picosecond acoustic microscopy for non-invasive characterization of buried interfaces," J. Mater. Res. **21**(5), 1204–1208 (2006).
28. J. C. D. Faria, P. Garnier, and A. Devos, "Non-destructive spatial characterization of buried interfaces in multilayer stacks via two color picosecond acoustics," Appl. Phys. Lett. **111**(24), 243105 (2017).
29. V. E. Gusev and P. Ruello, "Advances in applications of time-domain Brillouin scattering for nanoscale imaging," Appl. Phys. Rev. **5**(3), (2018).
30. O. Matsuda, M. C. Larciprete, R. Li Voti, and O. B. Wright, "Fundamentals of picosecond laser ultrasonics," Ultrasonics **56**, 3–20 (2015).
31. T. Saito, O. Matsuda, and O. B. Wright, "Picosecond acoustic phonon pulse generation in nickel and chromium," Phys. Rev. B **67**(20), 205421 (2003).
32. K. E. O'Hara, X. Hu, and D. G. Cahill, "Characterization of nanostructured metal films by picosecond acoustics and interferometry," J. Appl. Phys. **90**(9), 4852–4858 (2001).
33. C. Mechri, P. Ruello, and V. Gusev, "Confined coherent acoustic modes in a tubular nanoporous alumina film probed by picosecond acoustics methods," New J. Phys. **14**(2), 023048 (2012).
34. O. B. Wright and V. E. Gusev, "Ultrafast acoustic phonon generation in gold," Phys. B Condens. Matter **219**–**220**(1–4), 770–772 (1996).
35. C. Rossignol and B. Perrin, "Interferometric detection in picosecond ultrasonics for nondestructive testing of submicrometric opaque multilayered samples: TiN/AlCu/TiN/Ti/Si," IEEE Trans. Ultrason. Ferroelectr. Freq. Control **52**(8), 1354–1359 (2005).
36. A. Bartels, T. Dekorsy, H. Kurz, and K. Köhler, "Coherent Zone-Folded Longitudinal Acoustic Phonons in Semiconductor Superlattices: Excitation and Detection," Phys. Rev. Lett. **82**(5), 1044–1047 (1999).
37. Y. Ezzahri, S. Grauby, J. M. Rampnoux, H. Michel, G. Pernot, W. Claeys, S. Dilhaire, C. Rossignol, G. Zeng, and A. Shakouri, "Coherent phonons in Si/SiGe superlattices," Phys. Rev. B **75**(19), 195309 (2007).
38. B. C. Daly, T. B. Norris, J. Chen, and J. B. Khurgin, "Picosecond acoustic phonon pulse propagation in silicon," Phys. Rev. B **70**(21), 214307 (2004).
39. O. B. Wright and V. E. Gusev, "Acoustic generation in crystalline silicon with femtosecond optical pulses," Appl. Phys. Lett. **66**(10), 1190–1192 (1995).
40. D. B. Hondongwa, B. C. Daly, T. B. Norris, B. Yan, J. Yang, and S. Guha, "Ultrasonic attenuation in amorphous silicon at 50 and 100 GHz," Phys. Rev. B **83**(12), 121303 (2011).
41. Z. Vardeny, "Picosecond ultrasonics in thin films of conducting polymers," Synth. Met. **28**(3), D203–D208 (1989).
42. Y. C. Lee, K. C. Bretz, F. W. Wise, and W. Sachse, "Picosecond acoustic measurements of longitudinal

43. M. Hettich, K. Jacob, O. Ristow, M. Schubert, A. Bruchhausen, V. Gusev, and T. Dekorsy, "Viscoelastic properties and efficient acoustic damping in confined polymer nano-layers at GHz frequencies," Sci. Rep. **6**(1), 33471 (2016).
44. A. V. Akimov, E. S. K. Young, J. S. Sharp, V. Gusev, and A. J. Kent, "Coherent hypersonic closed-pipe organ like modes in supported polymer films," Appl. Phys. Lett. **99**(2), 021912 (2011).
45. P. Ruello, A. Ayouch, G. Vaudel, T. Pezeril, N. Delorme, S. Sato, K. Kimura, and V. E. Gusev, "Ultrafast acousto-plasmonics in gold nanoparticle superlattices," Phys. Rev. B - Condens. Matter Mater. Phys. **92**(17), 1–6 (2015).
46. C. L. Poyser, T. Czerniuk, A. Akimov, B. T. Diroll, E. A. Gaulding, A. S. Salasyuk, A. J. Kent, D. R. Yakovlev, M. Bayer, and C. B. Murray, "Coherent Acoustic Phonons in Colloidal Semiconductor Nanocrystal Superlattices," ACS Nano **10**(1), 1163–1169 (2016).
47. O. B. Wright and T. Hyoguchi, "Ultrafast vibration and laser acoustics in thin transparent films," Opt. Lett. **16**(19), 1529 (1991).
48. J. A. Rogers, A. A. Maznev, M. J. Banet, and K. A. Nelson, "Optical Generation and Characterization of Acoustic Waves in Thin Films: Fundamentals and Applications," Annu. Rev. Mater. Sci. **30**(1), 117–157 (2000).
49. N.-W. Pu and J. Bokor, "Study of Surface and Bulk Acoustic Phonon Excitations in Superlattices using Picosecond Ultrasonics," Phys. Rev. Lett. **91**(7), 076101 (2003).
50. M. F. P. Winter, A. Fainstein, B. Jusserand, B. Perrin, and A. Lemàtre, "Optimized optical generation and detection of superlattice acoustic phonons," Appl. Phys. Lett. **94**(10), 1–3 (2009).
51. K.-H. Lin, C.-F. Chang, C.-C. Pan, J.-I. Chyi, S. Keller, U. Mishra, S. P. DenBaars, and C.-K. Sun, "Characterizing the nanoacoustic superlattice in a phonon cavity using a piezoelectric single quantum well," Appl. Phys. Lett. **89**(14), 143103 (2006).
52. K. Mizoguchi, M. Hase, S. Nakashima, and M. Nakayama, "Observation of coherent folded acoustic phonons propagating in a GaAs/AlAs superlattice by two-color pump-probe spectroscopy," Phys. Rev. B **60**(11), 8262–8266 (1999).
53. H. Maris, M. Chand, W. Singhsomroje, and Z. Zannitto, "Studies of very small structures using picosecond ultrasonics," Wolrd Congr. Ultrason. 1105–1111 (2003).
54. D. Brick, V. Engemaier, Y. Guo, M. Grossmann, G. Li, D. Grimm, O. G. Schmidt, M. Schubert, V. E. Gusev, M. Hettich, and T. Dekorsy, "Interface Adhesion and Structural Characterization of Rolled-up GaAs/In0.2Ga0.8As Multilayer Tubes by Coherent Phonon Spectroscopy," Sci. Rep. **7**(1), 5385 (2017).
55. B. C. Daly, N. C. R. Holme, T. Buma, C. Branciard, T. B. Norris, D. M. Tennant, J. A. Taylor, J. E. Bower, and S. Pau, "Imaging nanostructures with coherent phonon pulses," Appl. Phys. Lett. **84**(25), 5180–5182 (2004).
56. P.-A. Mante, L. Belliard, and B. Perrin, "Acoustic phonons in nanowires probed by ultrafast pump-probe spectroscopy," Nanophotonics **7**(11), 1759–1780 (2018).
57. A. Amziane, L. Belliard, F. Decremps, and B. Perrin, "Ultrafast acoustic resonance spectroscopy of gold nanostructures: Towards a generation of tunable transverse waves," Phys. Rev. B **83**(1), 014102 (2011).
58. N. D. Lanzillotti-Kimura, A. Fainstein, A. Huynh, B. Perrin, B. Jusserand, A. Miard, and A. Lemaître, "Coherent Generation of Acoustic Phonons in an Optical Microcavity," Phys. Rev. Lett. **99**(21), 217405 (2007).
59. C. Jean, L. Belliard, T. W. Cornelius, O. Thomas, M. E. Toimil-Molares, M. Cassinelli, L. Becerra, and B. Perrin, "Direct Observation of Gigahertz Coherent Guided Acoustic Phonons in Free-Standing Single Copper Nanowires," J. Phys. Chem. Lett. **5**(23), 4100–4104 (2014).
60. C. Jean, L. Belliard, T. W. Cornelius, O. Thomas, Y. Pennec, M. Cassinelli, M. E. Toimil-Molares, and B. Perrin, "Spatiotemporal Imaging of the Acoustic Field Emitted by a Single Copper Nanowire," Nano Lett. **16**(10), 6592–6598 (2016).
61. H. Maris, "Picosecond ultrasonics," Sci. Am. **278**(1), 86–89 (1998).
62. C. Rossignol, N. Chigarev, M. Ducousso, B. Audoin, G. Forget, F. Guillemot, and M. C. Durrieu, "In Vitro picosecond ultrasonics in a single cell," Appl. Phys. Lett. **93**(12), 5–8 (2008).
63. T. Dehoux, M. A. Ghanem, O. F. Zouani, J. M. Rampnoux, Y. Guillet, S. Dilhaire, M. C. Durrieu, and B. Audoin, "All-optical broadband ultrasonography of single cells," Sci. Rep. **5**, 1–5 (2015).
64. L. Liu, L. Plawinski, M. C. Durrieu, and B. Audoin, "Label-free multi-parametric imaging of single cells: dual picosecond optoacoustic microscopy," J. Biophotonics **12**(8), 1–11 (2019).
65. F. Pérez-Cota, R. Fuentes-Domínguez, S. La Cavera, W. Hardiman, M. Yao, K. Setchfield, E. Moradi, S. Naznin, A. Wright, K. F. Webb, A. Huett, C. Friel, V. Sottile, H. M. Elsheikha, R. J. Smith, and M. Clark, "Picosecond ultrasonics for elasticity-based imaging and characterization of biological cells," J. Appl. Phys. **128**(16), (2020).
66. C. Thomsen, H. T. Grahn, H. J. Maris, and J. Tauc, "Surface generation and detection of phonons by picosecond light pulses," Phys. Rev. B **34**(6), 4129–4138 (1986).
67. G. L. Eesley, B. M. Clemens, and C. A. Paddock, "Generation and detection of picosecond acoustic pulses in thin metal films," Appl. Phys. Lett. **50**(12), 717–719 (1987).
68. W. S. Capinski and H. J. Maris, "Improved apparatus for picosecond pump-and-probe optical measurements," Rev. Sci. Instrum. **67**(8), 2720–2726 (1996).

69. K. J. Weingarten, M. J. W. Rodwell, H. K. Heinrich, B. H. Kolner, and D. M. Bloom, "Direct electro-optic sampling of GaAs integrated circuits," Electron. Lett. **21**(17), 765 (1985).
70. K. J. Weingarten, M. J. W. Rodwell, and D. M. Bloom, "Picosecond Optical Sampling of GaAs Integrated Circuits," IEEE J. Quantum Electron. **24**(2), 198–220 (1988).
71. P. A. Elzinga, R. J. Kneisler, F. E. Lytle, Y. Jiang, G. B. King, and N. M. Laurendeau, "Pump/probe method for fast analysis of visible spectral signatures utilizing asynchronous optical sampling," Appl. Opt. **26**(19), 4303 (1987).
72. O. Kliebisch, D. C. Heinecke, and T. Dekorsy, "Ultrafast time-domain spectroscopy system using 10 GHz asynchronous optical sampling with 100 kHz scan rate," Opt. Express **24**(26), 29930 (2016).
73. I. Coddington, N. Newbury, and W. Swann, "Dual-comb spectroscopy," Optica **3**(4), 414 (2016).
74. A. Bartels, F. Hudert, C. Janke, T. Dekorsy, and K. Köhler, "Femtosecond time-resolved optical pump-probe spectroscopy at kilohertz-scan-rates over nanosecond-time-delays without mechanical delay line," Appl. Phys. Lett. **88**(4), 1–3 (2006).
75. S. M. Link, A. Klenner, M. Mangold, C. A. Zaugg, M. Golling, B. W. Tilma, and U. Keller, "Dual-comb modelocked laser," Opt. Express **23**(5), 5521 (2015).
76. S. M. Link, A. Klenner, and U. Keller, "Dual-comb modelocked lasers: semiconductor saturable absorber mirror decouples noise stabilization," Opt. Express **24**(3), 1889 (2016).
77. S. M. Link, D. J. H. C. Maas, D. Waldburger, and U. Keller, "Dual-comb spectroscopy of water vapor with a free-running semiconductor disk laser," Science (80-. ). **356**(6343), 1164–1168 (2017).
78. J. Nürnberg, C. G. E. Alfieri, Z. Chen, D. Waldburger, N. Picqué, and U. Keller, "An unstabilized femtosecond semiconductor laser for dual-comb spectroscopy of acetylene," Opt. Express **27**(3), 3190 (2019).
79. J. Nürnberg, B. Willenberg, C. R. Phillips, and U. Keller, "Dual-comb ranging with frequency combs from single cavity free-running laser oscillators," Opt. Express **29**(16), 24910 (2021).
80. T. Ideguchi, A. Poisson, G. Guelachvili, N. Picqué, and T. W. Hänsch, "Adaptive real-time dual-comb spectroscopy," Nat. Commun. **5**, 3375 (2014).
81. R. Liao, H. Tian, W. Liu, R. Li, Y. Song, and M. Hu, "Dual-comb generation from a single laser source: Principles and spectroscopic applications towards mid-IR - A review," JPhys Photonics **2**(4), 0–17 (2020).
82. B. Willenberg, J. Pupeikis, L. M. Krüger, F. Koch, C. R. Phillips, and U. Keller, "Femtosecond dual-comb Yb:CaF2 laser from a single free-running polarization-multiplexed cavity for optical sampling applications," Opt. Express **28**(20), (2020).
83. R. Paschotta, H. R. Telle, and U. Keller, "Noise of Solid-State Lasers," in *Solid-State Lasers and Applications*, A. Sennaroglu, ed. (CRC Press, Taylor and Francis Group, LLC, 2007), pp. 473–510.
84. P. Kwee, B. Willke, and K. Danzmann, "Shot-noise-limited laser power stabilization with a high-power photodiode array," Opt. Lett. **34**(19), 2912 (2009).
85. S. Schilt, V. Dolgovskiy, N. Bucalovic, L. Tombez, M. C. Stumpf, G. Di Domenico, C. Schori, S. Pekarek, A. E. H. Oehler, T. Südmeyer, U. Keller, and P. Thomann, "Optical frequency comb with sub-radian CEO phase noise from a SESAM-modelocked 1.5-μm solid-state laser," Opt. InfoBase Conf. Pap. **19**(24), 635–639 (2011).
86. T. D. Shoji, W. Xie, K. L. Silverman, A. Feldman, T. Harvey, R. P. Mirin, and T. R. Schibli, "Ultra-low-noise monolithic mode-locked solid-state laser," Optica **3**(9), 995 (2016).
87. F. Saltarelli, I. J. Graumann, L. Lang, D. Bauer, C. R. Phillips, and U. Keller, "Power scaling of ultrafast oscillators: 350-W average-power sub-picosecond thin-disk laser," Opt. Express **27**(22), 31465 (2019).
88. L. M. Krüger, A. S. Mayer, Y. Okawachi, X. Ji, A. Klenner, A. R. Johnson, C. Langrock, M. M. Fejer, M. Lipson, A. L. Gaeta, V. J. Wittwer, T. Südmeyer, C. R. Phillips, and U. Keller, "Performance scaling of a 10-GHz solid-state laser enabling self-referenced CEO frequency detection without amplification," Opt. Express **28**(9), 12755 (2020).
89. K. Fritsch, J. Brons, M. Iandulskii, K. F. Mak, Z. Chen, N. Picqué, and O. Pronin, "Dual-comb thin-disk oscillator," (2020).
90. N. Modsching, J. Drs, P. Brochard, J. Fischer, S. Schilt, V. J. Wittwer, and T. Südmeyer, "High-power dual-comb thin-disk laser oscillator for fast high-resolution spectroscopy," Opt. Express **29**(10), 15104 (2021).
91. F. Druon, S. Ricaud, D. N. Papadopoulos, A. Pellegrina, P. Camy, J. L. Doualan, R. Moncorgé, A. Courjaud, E. Mottay, and P. Georges, "On Yb:CaF_2 and Yb:SrF_2: review of spectroscopic and thermal properties and their impact on femtosecond and high power laser performance [Invited]," Opt. Mater. Express **1**(3), 489 (2011).
92. I. Snetkov, A. Vyatkin, O. Palashov, and E. Khazanov, "Drastic reduction of thermally induced depolarization in CaF2 crystals with [111] orientation," Opt. Express **20**(12), 13357 (2012).
93. K. Genevrier, D. N. Papadopoulos, M. Besbes, P. Camy, J. L. Doualan, R. Moncorgé, P. Georges, and F. Druon, "Thermally-induced-anisotropy issues in oriented cubic laser crystals, the cryogenically cooled Yb:CaF2 case," Appl. Phys. B Lasers Opt. **124**(11), 1–12 (2018).
94. U. Keller, K. J. Weingarten, F. X. Kartner, D. Kopf, B. Braun, I. D. Jung, R. Fluck, C. Honninger, N. Matuschek, and J. Aus der Au, "Semiconductor saturable absorber mirrors (SESAM's) for femtosecond to nanosecond pulse generation in solid-state lasers," IEEE J. Sel. Top. Quantum Electron. **2**(3), 435–453 (1996).

95. G. Machinet, P. Sevillano, F. Guichard, R. Dubrasquet, P. Camy, J. Doualan, and R. Moncorgé, "Kerr-lens mode-locked Yb : CaF2 oscillator," **38**(20), 4008–4010 (2013).
96. F. X. Kartner, I. D. Jung, and U. Keller, "Soliton mode-locking with saturable absorbers," IEEE J. Sel. Top. Quantum Electron. **2**(3), 540–556 (1996).
97. K.-H. Lin, C.-T. Yu, Y.-C. Wen, and C.-K. Sun, "Generation of picosecond acoustic pulses using a p-n junction with piezoelectric effects," Appl. Phys. Lett. **86**(9), 093110 (2005).
98. K.-H. Lin, C.-M. Lai, C.-C. Pan, J.-I. Chyi, J.-W. Shi, S.-Z. Sun, C.-F. Chang, and C.-K. Sun, "Spatial manipulation of nanoacoustic waves with nanoscale spot sizes," Nat. Nanotechnol. **2**(11), 704–708 (2007).
99. C. K. Sun, J. C. Liang, and X. Y. Yu, "Coherent Acoustic Phonon Oscillations in Semiconductor Multiple Quantum Wells with Piezoelectric Fields," Phys. Rev. Lett. **84**(1), 179–182 (2000).
100. Y.-C. Wen, L.-C. Chou, H.-H. Lin, V. Gusev, K.-H. Lin, and C.-K. Sun, "Efficient generation of coherent acoustic phonons in (111) InGaAs⁄GaAs multiple quantum wells through piezoelectric effects," Appl. Phys. Lett. **90**(17), 172102 (2007).
101. P.-A. Mante, Y.-R. Huang, S.-C. Yang, T.-M. Liu, A. A. Maznev, J.-K. Sheu, and C.-K. Sun, "THz acoustic phonon spectroscopy and nanoscopy by using piezoelectric semiconductor heterostructures," Ultrasonics **56**, 52–65 (2015).
102. G. Arregui, N. D. Lanzillotti-Kimura, C. M. Sotomayor-Torres, and P. D. García, "Anderson Photon-Phonon Colocalization in Certain Random Superlattices," Phys. Rev. Lett. **122**(4), 043903 (2019).
103. G. Arregui, O. Ortíz, M. Esmann, C. M. Sotomayor-Torres, C. Gomez-Carbonell, O. Mauguin, B. Perrin, A. Lemaître, P. D. García, and N. D. Lanzillotti-Kimura, "Coherent generation and detection of acoustic phonons in topological nanocavities," APL Photonics **4**(3), 030805 (2019).
104. F. R. Lamberti, Q. Yao, L. Lanco, D. T. Nguyen, M. Esmann, A. Fainstein, P. Sesin, S. Anguiano, V. Villafañe, A. Bruchhausen, P. Senellart, I. Favero, and N. D. Lanzillotti-Kimura, "Optomechanical properties of GaAs/AlAs micropillar resonators operating in the 18 GHz range," Opt. Express **25**(20), 24437 (2017).
105. J. Ravichandran, A. K. Yadav, R. Cheaito, P. B. Rossen, A. Soukiassian, S. J. Suresha, J. C. Duda, B. M. Foley, C. H. Lee, Y. Zhu, A. W. Lichtenberger, J. E. Moore, D. A. Muller, D. G. Schlom, P. E. Hopkins, A. Majumdar, R. Ramesh, and M. A. Zurbuchen, "Crossover from incoherent to coherent phonon scattering in epitaxial oxide superlattices," Nat. Mater. **13**(2), 168–172 (2014).
106. C. Li, V. Gusev, T. Dekorsy, and M. Hettich, "All optical control of comb-like coherent acoustic phonons in multiple quantum well structures through double-pump-pulse pump-probe experiments," Opt. Express **27**(13), 18706 (2019).
107. C. Li, V. Gusev, E. Dimakis, T. Dekorsy, and M. Hettich, "Broadband photo-excited coherent acoustic frequency combs and mini-Brillouin-zone modes in a MQW-SESAM structure," Appl. Sci. **9**(2), 1–22 (2019).
108. J. Shah, *Ultrafast Spectroscopy of Semiconductors and Semiconductor Nanostructures*, Springer Series in Solid-State Sciences (Springer Berlin Heidelberg, 1999), **115**.
109. C. Li, V. Gusev, T. Dekorsy, and M. Hettich, "All optical control of comb-like coherent acoustic phonons in multiple quantum well structures through double-pump-pulse pump-probe experiments," Opt. Express **27**(13), 18706 (2019).
110. C. Li, N. Krauß, G. Schäfer, L. Ebner, O. Kliebisch, J. Schmidt, S. Winnerl, M. Hettich, and T. Dekorsy, "High-speed asynchronous optical sampling based on GHz Yb:KYW oscillators," Opt. Express **25**(8), 9204 (2017).
111. H. Neumann, "S. Adachi (ed.). Properties of Aluminium Gallium Arsenide. (EMIS Datareviews Series No. 7). INSPEC; The Institution of Electrical Engineers, London 1993. 325 S., 58 Abb., 93 Tab. ISBN 0 85296 558 3, L 95," Cryst. Res. Technol. **28**(6), 866–866 (1993).
112. M. Levinshtein, S. Rumyantsev, and M. Shur, *Handbook Series on Semiconductor Parameters* (WORLD SCIENTIFIC, 1996).
113. B. Bonello, B. Perrin, and C. Rossignol, "Photothermal properties of bulk and layered materials by the picosecond acoustics technique," J. Appl. Phys. **83**(6), 3081–3088 (1998).
114. O. L. Anderson, "Determination and Some Uses of Isotropic Elastic Constants of Polycrystalline Aggregates Using Single-Crystal Data," in *Physical Acoustics* (ACADEMIC PRESS INC., 1965), **3**(1928), pp. 43–95.
115. C. Wolff, M. J. A. Smith, B. Stiller, and C. G. Poulton, "Brillouin scattering—theory and experiment: tutorial," J. Opt. Soc. Am. B **38**(4), 1243 (2021).
116. A. Bartels, R. Cerna, C. Kistner, A. Thoma, F. Hudert, C. Janke, and T. Dekorsy, "Ultrafast time-domain spectroscopy based on high-speed asynchronous optical sampling," Rev. Sci. Instrum. **78**(3), (2007).